\documentclass[twocolumn,showpacs,preprintnumbers,amsmath,amssymb]{revtex4}

\usepackage[dvips]{graphicx}
\usepackage{dcolumn}
\usepackage{bm}
\usepackage{multirow}
\usepackage{array}
\usepackage{mathtools}
\usepackage{subfig}
\newcolumntype{P}[1]{>{\centering\arraybackslash}p{#1}}
\usepackage[utf8]{inputenc}
\usepackage[LGR,T1]{fontenc}

\begin{document}

\preprint{APS/123-QED}

\title{A Three State Opinion Formation Model for Financial Markets}

\author{Bernardo J. Zubillaga}
\affiliation{%
Boston University, Center for Polymer Studies and Department of Physics, Boston, 02115, USA
}%

\author{Andr\'e L. M. Vilela}%
\affiliation{%
Boston University, Center for Polymer Studies and Department of Physics, Boston, 02115, USA
}%
\affiliation{%
Universidade de Pernambuco, Recife, Pernambuco, 50100-010, Brazil.
}%

\author{Chao Wang}%
 \email{chaowanghn@vip.163.com}
\affiliation{%
Beijing University of Technology, School of Economics and Management, Beijing, 100124, China.
}%
\affiliation{%
Boston University, Center for Polymer Studies and Department of Physics, Boston, 02115, USA.
}%

\author{Kenric P. Nelson}%
\affiliation{%
Boston University, Electrical and Computer Engineering, Boston, MA 02215, USA.
}%

\author{H. Eugene Stanley}%
\affiliation{%
Boston University, Center for Polymer Studies and Department of Physics, Boston, 02115, USA.
}%

\date{\today}

\begin{abstract}
We propose a three-state microscopic opinion formation model for the purpose of simulating the dynamics of financial markets.
In order to mimic the heterogeneous composition of the mass of investors in a market, 
the agent-based model considers two different types of traders:
noise traders and contrarians. Agents are represented as nodes in a network of interactions and they can assume any of three distinct possible
states (e.g. buy, sell or remain inactive). The time evolution of the state of an agent
is dictated by probabilistic dynamics that include both local and global influences.
A noise trader is subject to local interactions, tending to assume the majority state of its
nearest neighbors, whilst a contrarian is subject to a global interaction with the behavior of the market as a whole,
tending to assume the state of the global minority of the market.
The model exhibits the typical qualitative and quantitative features of real financial time series, including distributions of returns with heavy tails, volatility clustering and long-time memory for the absolute
values of the returns. The distributions of returns are fitted by means of coupled Gaussian distributions,
quantitatively revealing transitions between leptokurtic, mesokurtic and platykurtic regimes in terms of 
a non-linear statistical coupling which describes the complexity of the system.

\end{abstract}

\maketitle

\section{INTRODUCTION}
Over the course of the past few decades, there has been widespread interest in the implementation of tools and methods of statistical mechanics
for the purpose of investigating the behavior and dynamics of social and economic systems 
\cite{Mantegna2000,Bouchaud2003,Voit,Bornholdt,Kaizoji2002,Takaishi2005,WeronWeron2002,KKH2002,BLT2005,Vilela,DeOliveira,Brunstein1999,Tome2002,Melo2010, Vilela2018, Sornette2006}.
Statistical physics studies the complex behavior of macroscopic physical systems in terms of the basic interactions between their numerous
fundamental components. A hallmark of complexity is the emergence of collective or cooperative effects which are not reducible to the behavior
of any individual component. Financial markets are complex systems composed of millions of investors worldwide, partaking of commercial
activities, interacting amongst themselves for the purposes of buying and selling financial assets such as stock, bonds, options and commodities.
The convergence of individual decisions between different financial agents has repercussions on the behavior of macroscopic observables such as
the time series of prices of such assets. 

Therefore, one of the fundamental questions on financial systems is the formulation of simple models of microscopic dynamics capable of reproducing
the essential quantitative and qualitative statistical features of real financial time series. This task suggests an examination of the
possible underlying processes and dynamics of opinion formation in a market, which ultimately drive the financial decisions of purchase or sale of
assets. This process involves not merely financial considerations, but also psychological, emotional and social factors.

The study of microscopic models of opinion formation, aided by computer simulations, may shed light on the essential interaction laws between
agents in a financial market, which can give rise to the salient behavior of macroscopic financial observables 
as a consequence of those microscopic interactions, such as heavy-tailed return distributions and volatility clustering
\cite{Vilela,Bornholdt,Kaizoji2002,WeronWeron2002,KKH2002,BLT2005,Voit}.

Microscopic models of opinion formation do not pretend to mimic human opinion formation dynamics in an absolutely rigorous fashion.
Rather, they attempt to reduce such processes to very basic interaction laws with very few parameters in an attempt to understand 
possible candidates for the fundamental mechanisms at play that give rise to such macroscopic complexity.

Ising-like systems have been proposed in the past as frameworks for agent-based models for financial systems:
One such model, proposed by Bornholdt, considers the Ising model with global interactions that couple the agents/spins to the global magnetization of the system
as well as local interactions between each agent and its nearest neighbors \cite{Bornholdt}.
Metastable states ensue in this model below its critical temperature as a consequence of the competition between local and global
interactions. 
Sznajd-Weron and Weron proposed a microscopic model of price formation, in which a one dimensional chain of spins is endowed with dynamical rules for outward flow of information and a heterogenous composition of agents \cite{WeronWeron2002}. 
Other examples include the models by Krawiecki, Holyst and Helbing, whose Ising-like mean field model implements heat bath dynamics in such a way that 
the interactions between the agents change randomly over time \cite{KKH2002}.
A similar model implementing heat bath dynamics with stochastic and time-varying interactions considers only nearest neighbor interactions on
non-trivial topologies of Barab\'asi-Albert networks \cite{BLT2005}.
These simple models are capable of reproducing principal features of real financial time series such as long-term correlations and scaling.

The majority-vote model is another such widely studied agent-based model in statistical mechanics \cite{DeOliveira,Brunstein1999,Tome2002, Melo2010, Vilela2018}.
It was formulated in order to study the dynamics of opinion formation in a society.
The three-state version of the model assumes that individuals are nodes placed in a social network in any one of three possible states/opinions, 
interacting with their nearest neighbors, exerting influence on them and being influenced by them in return.
In this model, with probability $1-q$, the agent will agree with the majority state of its neighbors or to dissent from it with probability $q$, also known as the noise parameter. The three-state majority-vote model (MVM3) exhibits a second order phase transition in a square lattice network for $q_c \approx 0.118$.
\cite{Brunstein1999,Tome2002,Melo2010}

The dynamics of financial markets ensues not only from rational action on the part of investors but also from emotional behaviors, which are 
a reflection of the rich social psychology of the influences and interactions between agents in a network of financial investors.
An example of such a phenomenon is herding, whereby individuals tend to follow the opinion or behavior of their
neighbors \cite{Raafat2009,Hong2005}. Herding behavior has been suggested to play an essential role in human behavior and in the animal kingdom,
in situations that range from collective behaviors of flocks of birds and school of fish to riots, strikes, sporting events and opinion formation
(where often coherence in social imitation manifests as informational cascades \cite{Bikhchandani1992}).
In the context of financial markets, the field of behavioral finance has identified herding as a key to the understanding of the
collective irrationality of investors \cite{Shiller2015}. It is carried out by so-called noise traders.
They typically follow the trends of their neighbors and have a propensity to overreact to the arrival of news when buying or selling.
 
Other agents in a market seem to follow the trends of the global minority as an investment strategy.
Thus they tend to buy when noise traders drive the prices down and they tend to sell when noise traders drive the prices up.
We shall refer to these agents as contrarians \cite{Voit,Bornholdt,Kaizoji2002,Takaishi2005,Lux1999,Lux2000,DeLong1990,Day1990}. 
Rational decision-making on their part tends to drive prices towards the values suggested by the analysis of the
fundamentals of an asset such as expected profits of the company, expected interests rates, dividends, future plans of the company, etc.

Based on the MVM3, we propose a generalization of the opinion formation model for financial markets introduced by Vilela et al.
into a three-state model, including a state where an investor remains inactive \cite{Vilela}. In this work, we shall classify investors in a market as either noise traders or contrarians, 
a simple scheme that models a heterogeneous composition of the mass of investors. 
Noise traders follow the local minority via herding behavior, while contrarians follow the global minority, consistent with fundamental analysis.
Agents are represented as nodes in a square lattice and they take any one of three possible states, with their opinion being influenced by the states of their
neighbors or the global state of the system. 
In this way, the social psychology of opinion formation dynamics serves as a basis for our microscopic market model.

This paper is organized as follows. In Section \ref{sec:model} we describe the model and dynamics of the agents.
In Section \ref{sec:results} we present the results of the simulations and numerical calculations with their corresponding discussions,
followed by our concluding remarks in Section \ref{sec:conclusion}.

\section{THE MODEL}
\label{sec:model}
Financial agents are represented by nodes in a regular 2D lattice network of size $N = L\times L$ with periodic boundary conditions.
Each agent assumes any one of three available states $s\in\{1,2,3\}$ at any instant of time.
In a financial context, such states could represent, for example, a desire to purchase a unit of an asset, 
to sell a unit of an asset or to remain neutral/inactive.

In order to mimic the heterogeneous composition of agents in a real-world financial market,
we consider two types of agents in this model: noise traders and contrarians.
Let $f$ represent the fraction of contrarian agents in the model and $1-f$ is the fraction of noise traders.

\subsection{Noise traders}
The state of a noise trader is updated according to the following probabilistic prescription,
in accordance with the majority-vote model with 3 states. The behavior of a noise trader is given by its tendency 
to follow the local majority, i.e., by its tendency to agree with the state of the majority of its nearest neighbors with probability $1-q$, or to dissent from it with probability $q$. Let $i$ be a noise trader. In the event of a single local majority state, agent $i$ will adopt it
with probability $1-q$ and each of the two local minority states will be adopted by $i$ with probability 
$q/2$. In the event of a tie between two local majority states, agent $i$ shall assume
any of those two states with probability $(1-q)/2$ each and the state of the local minority with
probability $q$.

Let $k_{i, s}$ represent the number of nearest neighbors of noise trader $i$ that find
themselves in state $s\in\{1,2,3\}$, with $k_{i, 1}+k_{i, 2}+k_{i, 3}=4$ for square lattice networks.
The aforementioned rules for the update of the state of noise trader $i$ can be summarized thus: 
\begin{equation}
    \label{eq:noisetraders}
    \begin{aligned}
    & P(1|k_{i,1}>k_{i,2}; k_{i,3})= 1-q, \\
    & P(1|k_{i,1}=k_{i,2}>k_{i,3})=(1-q)/2,\\    
    & P(1|k_{i,1}<k_{i,2}=k_{i,3})=q,\\
    & P(1|k_{i,1} ; k_{i,2}<k_{i,3})=q/2.\\
    \end{aligned}
\end{equation}
The probabilities for the remaining two states (2 and 3) follow easily from the
symmetry operations of the $C_{3\nu}$ group.
It is worth noticing that the condition 
$P(1|\{k_i\})+P(2|\{k_i\})+P(3|\{k_i\})=1$ holds for any configuration $\{k_i\}\equiv\{k_{i,1};k_{i,2};k_{i,3}\}$
of the local neighbors, as it should for the update probabilities
to be conserved.

For the model to make sense, one must have $1-q > q/2$, i.e.,
the probability for a noise trader to align with the local
majority has to be greater than the probability of aligning
with any of the two local minorities. This condition implies
that the model is defined for values of the noise parameter
in the range $0\leq q\leq 2/3$.

\subsection{Contrarians}

The behavior of contrarian agent is defined by its tendency
to follow the global minority, i.e., by its tendency to agree with
the state of the minority out of all $N$ traders in the system with
probability $1-q$, or to dissent from it with probability $q$.
This strategy is a non-local interaction between the agent and the market as a whole. The probabilistic rules of update for the state $s$ of contrarian 
agent $j$ are described as follows. Let $j$ be a contrarian agent and in the event of a single global minority state, agent $j$ will adopt it
with probability $1-q$ and each of the two global majority states will be adopted by $j$ with probability 
$q/2$. In the event of a tie between two global minority states, agent $j$ shall assume
any of those two states with probability $(1-q)/2$ each and the state of the global majority with
probability $q$. Finally, in the event of a three-way tie, agent $j$ will assume any state with 
probability $1/3$.

Let $N_{s}$ represent the total number of traders in the market that find
themselves in state $s\in\{1,2,3\}$, where $N_1+N_2+N_3=N$.
In this case, the rules for the update of the state of contrarian agent $j$ can be summarized as follows: 
\begin{equation}
    \label{eq:contrarians}
    \begin{aligned}
    & P(1|N_{1}<N_{2}; N_{3})= 1-q, \\
    & P(1|N_{1}=N_{2}<N_{3})=(1-q)/2,\\    
    & P(1|N_{1}>N_{2}=N_{3})=q,\\
    & P(1|N_{1};N_{2}>N_{3})=q/2,\\
    & P(1|N_{1}=N_{2}=N_{3})=1/3.
    \end{aligned}
\end{equation}

The probabilities for the remaining two states (2 and 3) follow easily from the
symmetry operations of the $C_{3\nu}$ group.
It is worth noticing that the condition 
$P(1|\{N\})+P(2|\{N\})+P(3|\{N\})=1$ also holds for any global state configuration $\{N\}\equiv \{N_1,N_2,N_3\}$.

\subsection{The Order Parameter}

For the purposes of this model, we shall adopt the magnitude of the magnetization, defined in 
analogy to the three-state Potts model, as an order parameter for this system.
The magnetization \textbf{M} is to be taken as a vector with components:
$M_s$ for $s\in\{1,2,3\}$.
Its magnitude $M$ is thus given by $M=\sqrt{M_1^2 +M_2^2 + M_3^2}$ and
its components are calculated according to:
\begin{equation}
M_s = \sqrt{\frac{3}{2}} \left[\frac{N_{s}}{N} - \frac{1}{3}\right].
\end{equation}
With this definition, it can be shown that the components of the magnetization
are not independent of one another. In fact, it holds that $M_{1}+M_{2}+M_{3}=0$.

\section{NUMERICAL RESULTS AND DISCUSSION}
\label{sec:results}

We now present the numerical results of Monte Carlo simulations of the stochastic dynamics introduced in the 
previous section on a square lattice of size $N=100\times100$ with periodic boundary conditions. In each simulation, each node is randomly assigned to be a contrarian or a noise trader with probability $f$ or $1-f$, respectively. Time is measured in units of Monte Carlo Steps (MCS).
In one MCS, a total of $N$ attempts are made to change the states of the agents.
In each attempt, a node is randomly selected in the network and its state is updated according to Eqs. (\ref{eq:noisetraders}), if the node is a noise trader, or Eqs. (\ref{eq:contrarians}), if the selected node is a contrarian.

The initial state of the system is random, assigning to each agent any one of the three available states with equal probability. In each simulation, a total of $20000$ MCSs are performed, of which the first $10000$ MCS are discarded as thermalization time of the
system. A total of $100$ Monte Carlo simulations were performed for every set of parameters $(q,f)$ considered. Therefore, effectively, a total of $10^6$ MCSs were recorded for each pair of parameters $(q,f)$ from all runs. In the absence of contrarians ($f=0$), it is clear that the dynamics reduces to the MVM3, which, on a $2D$ square lattice, exhibits a critical point at $q_c \approx 0.118$ in the thermodynamic limit ($N \to \infty$). For $q<q_c$, the system
exhibits an ordered phase with the presence of large clusters of agents that share the same state, which results in spontaneous
magnetization. For $q>q_c$, thermal fluctuations fully destroy the order and the order parameter, $M$, vanishes in the thermodynamic limit.

\begin{figure*}[ht]
\vspace{0.3cm}
    \centering
    \subfloat[Subfigure 1 list of figures text][]
        {
        \includegraphics[width=0.2\textwidth]{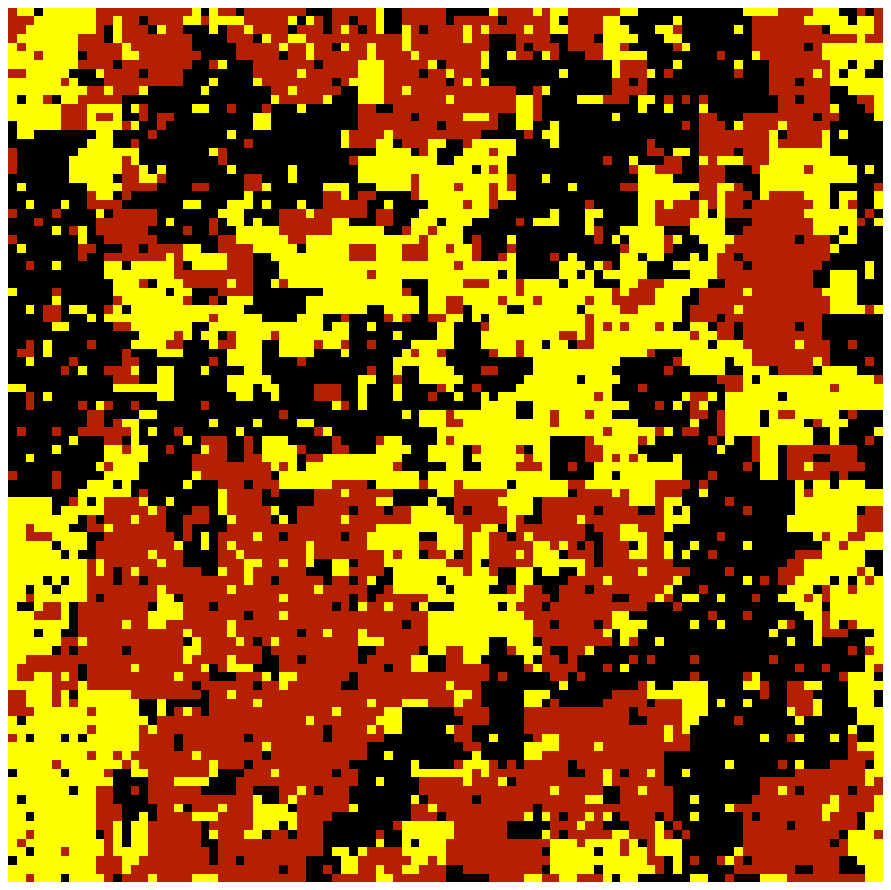}
        \label{fig:subfig1}
        }
    \subfloat[Subfigure 2 list of figures text][]
        {
        \includegraphics[width=0.2\textwidth]{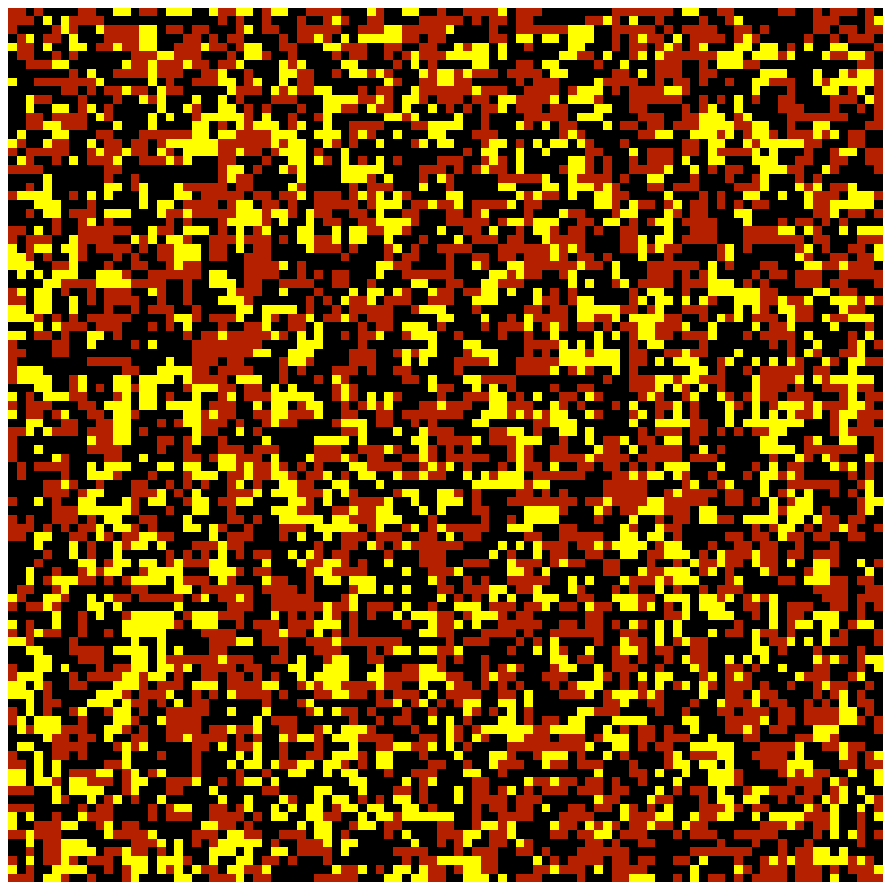}
        \label{fig:subfig2}
        }
    \subfloat[Subfigure 3 list of figures text][]
        {
        \includegraphics[width=0.2\textwidth]{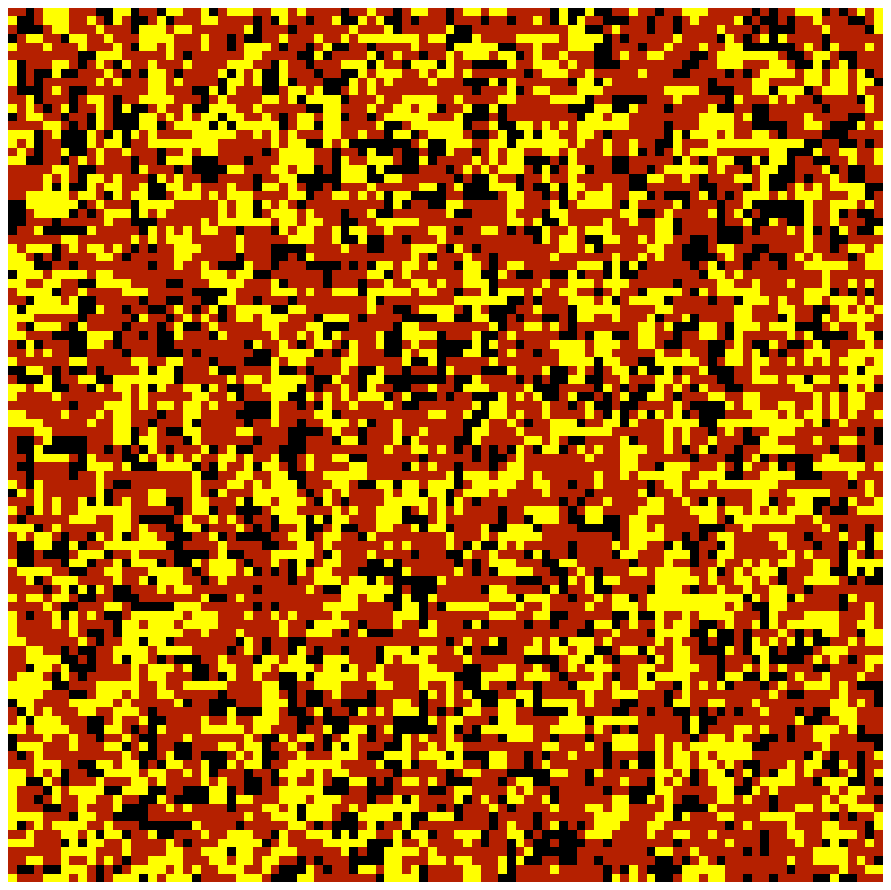}
        \label{fig:subfig3}
        }
    \subfloat[Subfigure 4 list of figures text][]
        {
        \includegraphics[width=0.2\textwidth]{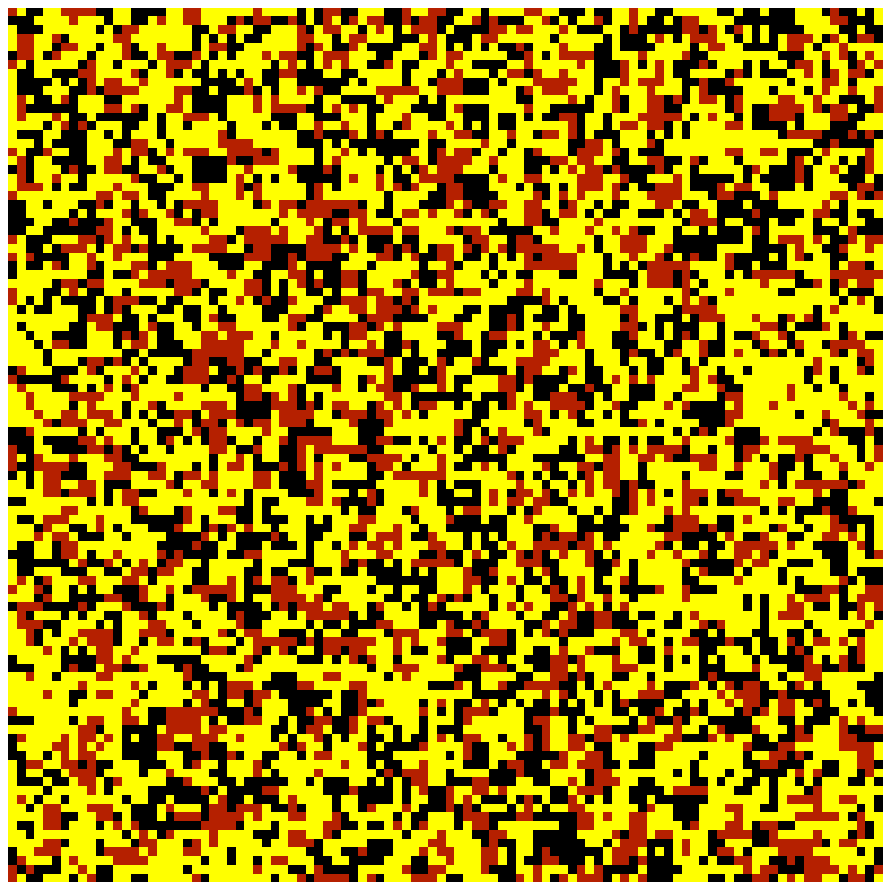}
        \label{fig:subfig4}
        }
    \caption{Snapshots of Monte Carlo simulations of the model on a square network of agents with periodic boundary conditions.
    Each square represents an agent and the three available states are depicted in black, red and yellow.
    The noise parameter for these simulations is $q=0.12$ and the fractions of contrarians are equal to $f=0.0$ for (a) and $f=0.5$ for (b), (c) and (d) where we have black, red and yellow as the majority state, respectively.}
    \label{fig:snapshots}
\end{figure*}

In Fig. \ref{fig:snapshots} we present a visual representation of individual snapshots of the states of the system for two concentration of
contrarians $f=0.0$ and $0.5$ to get a sense of the microscopic behavior of the system. In the absence of contrarians [Fig. \ref{fig:snapshots}(a)], the system clearly exhibits clusters of ordered opinions
near criticality ($q \sim q_c(f = 0) \sim 0.118$). The introduction of contrarians to the system induces the destruction of clusters of local order [Fig. \ref{fig:snapshots}(b), \ref{fig:snapshots}(c) and \ref{fig:snapshots}(d)]. For high values of $f$ the global
interactions of contrarians with may induce an order of their own as contrarians collectively seek the state of the
global minority en masse. This enables oscillatory behaviors in the populations of the states of the system, with cyclical regimes driven by a tendency of contrarians to flee from a global majority state into a global minority state. In Fig. \ref{fig:snapshots}(b) we see a metastable state where the state $s = 1$ (black) is the predominant state of the system. By running it further we check that the majority state can also oscillate into majority state $s = 2$ (red) and $s = 3$ (yellow). 

\begin{figure*}[ht]
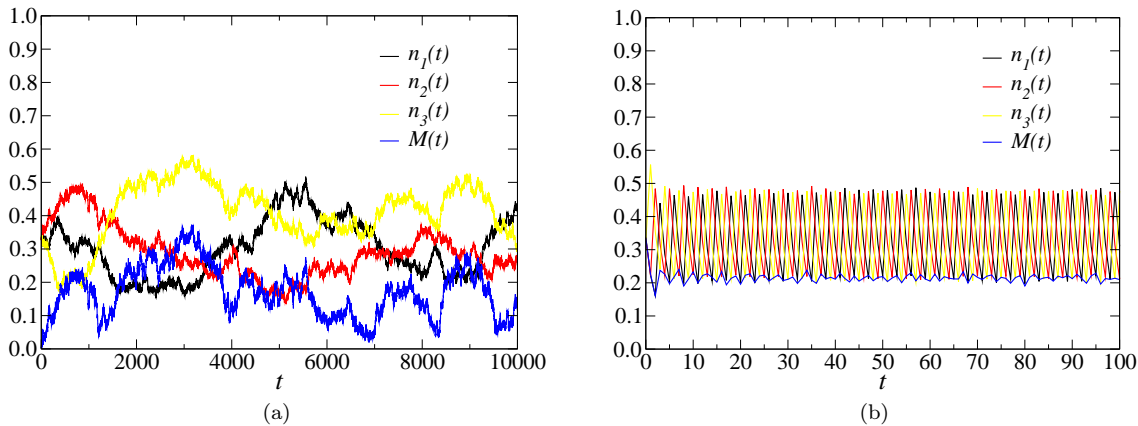

\vspace{0.3cm}
    \centering
    \subfloat[Subfigure 1 list of figures text][]
        {
        \includegraphics[width=0.4\textwidth]{censusf000.eps}
        \label{fig:subfig1}
        }
        \qquad
    \subfloat[Subfigure 2 list of figures text][]
        {
        \includegraphics[width=0.39\textwidth]{censusf050.eps}
        \label{fig:subfig2}
        }
    \caption{Time evolution of the fraction of agents in each of the three states $n_s$ with $s = 1$ (black), $2$ (red) and $3$ (yellow) for $q=0.12$.
   The magnitude of the order parameter $M$ of the system is shown in blue.
    The fraction of contrarians are: $f=0.00$ and $f=0.50$ for (a) and (b), respectively.}
    \label{fig:census}
\end{figure*}

By calculating the fraction of agents in each state $n_s(t) = N_s (t)/N$ and the magnitude of the order parameter $M(t)$ as a function of time, we find the typical time evolution of the relative populations shown in Figure \ref{fig:census}. 
The oscillatory pattern of the majority can be verified in the behavior of $n_s (t)$ shown in Fig. \ref{fig:census}(b).

Recalling that the fluctuations of the order parameter diverge at the critical point in the thermodynamic limit, 
the destruction of local clusters of ordered opinion by increasing the number of contrarians drives the system away from criticality, thereby
reducing the magnitude of the relative time fluctuations of the magnetization, which, in the context of a financial 
interpretation of this system, may suggest an increase of market stability.

At an instant of time $t$, we will relate the time variations of the magnitude of the magnetization $M(t)$ of this model to the 
logarithmic returns $r(t)$ of a financial asset (such as a stock or an index fund, for example), understanding each of the three possible 
opinions or states $s\in\{1,2,3\}$ of the agents as intentions to buy a unit of that asset, to sell a unit of that asset or to do 
neither or remain inactive, for example. We define the logarithmic return at time $t$ in terms of the magnetization
\begin{equation}
    \label{eq:return}
    r(t) \equiv \ln[M(t)] - \ln[M(t-1)]
\end{equation}
whereby $M(t)$, being a positive-definite quantity, can be interpreted as a measure of the price of the asset. We shall also define the volatility of the asset as a measure of the dispersion of the relative variations of the prices locally in time:
\begin{equation}
    v(t)\equiv|r(t)|,
\end{equation}
i.e., we define the volatility to be the absolute value of the logarithmic return. We now study the statistical properties of the time series of returns generated by the model.

\begin{figure}[h]
\vspace{0.3cm}
    \centering
    \includegraphics[width=0.4\textwidth]{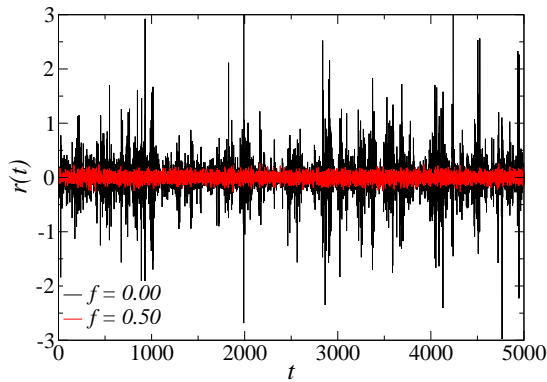} 
    \caption{Time series of logarithmic returns for a noise parameter $q=0.16$ and concentrations of contrarians given by $f=0.00$
    and $f=0.50$. In the absence of contrarians, the time series clearly exhibits volatility clustering. 
    }
   \label{fig:sample}
\end{figure}

In Fig. \ref{fig:sample}, we present an example of time series of returns produced by this model for $q=0.16$ and for two different
concentrations of contrarians: $f=0.00$ and $f=0.50$. 
As shown in the figure, the presence of contrarians serves to noticeably decrease the size of the 
fluctuations in the returns. 

Qualitatively, it is also visually clear that the time series in the absence of
contrarians exhibits \textit{volatility clustering}, a well known feature of real time series of returns first identified by
Mandelbrot, whereby large (small) fluctuations tend to follow by large (small) fluctuations. This suggests that a measure of nonlinear correlations, i.e., correlations of volatility, should exhibit this effect of memory
and structure of the time series \cite{Mandelbrot1963,Voit,Mantegna2000}. Volatility clusters are an essential feature of real financial time series, 
particularly prominent during times of bubbles and financial crashes, often a reflection of the collective mechanism of herding behavior 
driving the market, which has been shown to contradict the basic assumptions of perfectly rational behavior of investors
in financial systems \cite{Shiller2015}.

In order to quantify the effect of volatility clustering on the long-term memory of the volatility, we define its autocorrelation
function thus:
\begin{equation}
    \rho(\tau)\equiv\frac{\sum_{t=1}^{T-\tau}[|r(t)|-\langle|r|\rangle][|r(t+\tau)|-\langle|r|\rangle]}{\sum_{t=1}^{T}[|r(t)|-\langle|r|\rangle]^2}
\end{equation}

\begin{figure}[ht]
\vspace{0.3cm}
    \centering
    \includegraphics[width=0.4\textwidth]{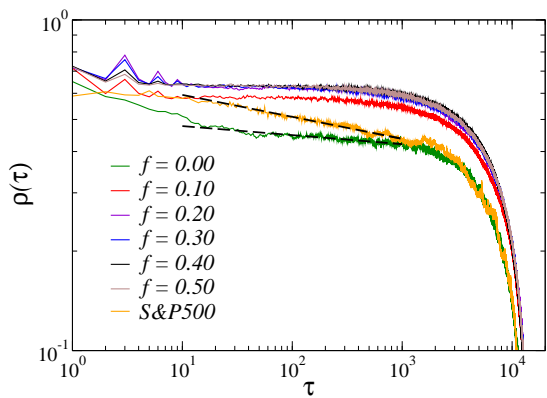} 
    \caption{Autocorrelation function of volatility for a noise parameter $q=0.16$ and various concentrations of contrarians: $f=0.00,0.10,0.20,0.30,0.40,0.50$. Also shown is the autocorrelation function of the daily volatility of the closing values of the  S\&P500 from to Dec 12, 1958 to June 25, 2018, for a total of $15000$ days. The dashed lines correspond to fits to power laws.}
   \label{fig:acfvol}
\end{figure}

Fig. \ref{fig:acfvol} exhibits the behavior of the autocorrelation function of the absolute values of the log-returns 
for different densities of contrarians, a value of $q=0.16$, a simulation of time $T=15000$.
Also, for comparison purposes, we show the autocorrelation function of daily log returns of the closing values of the S\&P500
from to Dec 12, 1958 to June 25, 2018, for a total of $15000$ days.
As is visually clear from the plots, there is qualitative agreement between real data and the simulations,
the curve of the S\&P500 lying somewhere between that of $f=0.00$ and $f=0.10$.

As in the 2 state model proposed by Vilela et al \cite{Vilela} and the Bornholdt model \cite{Bornholdt,Takaishi2005}, 
we observe that the time series of the volatility exhibit long-range memory, in agreement with the main features of non-stationary, real time series.
In effect, as has been observed in empirically in real financial data since Mandelbrot's initial observations, the autocorrelation of the
volatility is known to decrease typically as a power law, suggesting a lack of a characteristic time scale for the time series of absolute values
of returns.

It is obvious from Fig. \ref{fig:acfvol} that a deviation from this scale invariant regime eventually sets in after a time lag of 
about $\tau \approx 10^3$, as evidenced by the stretched exponential decay that follows. 
This decay is a reflection of the finite sizes of the data samples presented for the time series of both the simulations and the S\&P500.

As a manner of comparison, we present the fits (in dashed lines) of the autocorrelation function for $f=0.00$ and the 
empirical data of the S\&P500 to power laws, i.e., $\rho(\tau)\sim\tau^{-\eta}$, over the same time periods.
The exponents of the fits are $\eta=-0.0273\pm 0.0005$ for the simulation with $f=0.00$
and $\eta=-0.0664\pm 0.0005$ for the daily volatility of the S\&P500, which shows agreement in the order of the rate of decay of correlations
between simulations and real data.\\

\begin{figure}[h!]
\vspace{0.3cm}
    \centering
    \includegraphics[width=0.4\textwidth]{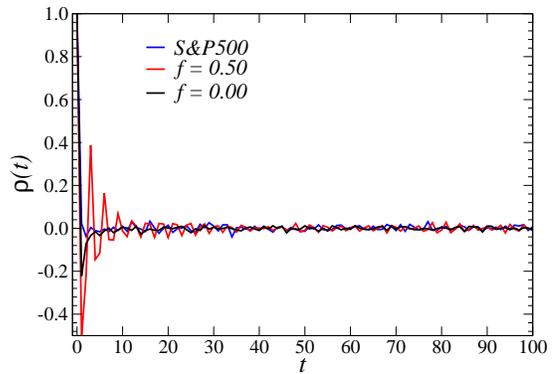} 
    \caption{Autocorrelation function of the logarithmic returns for a noise parameter $q=0.16$ and two concentrations of contrarians:
    $f=0.00$ and $0.50$. 
    Also shown is the autocorrelation function of the daily returns of the closing values of the S\&P500 from to Dec 12, 1958 to March 22, 1959 for a total of $100$ days.}
   \label{fig:acfr}
\end{figure}

A calculation of the autocorrelation function for the time series of returns produced by the model suggests that the returns
are essentially uncorrelated, a feature that is consistent with the \textit{efficient market hypothesis} \cite{Voit}.
This is shown in Fig. \ref{fig:acfr},
where the autocorrelation function of the logarithmic returns are plotted for simulations of the dynamics with
two different concentrations of contrarians, $f=0.00$ and $f=0.50$ and the autocorrelation function of the daily returns of the 
S\&P500 is also shown for comparison.
The agreement between real financial data and the simulation in the absence of contrarians is very clear.
In the case of $f=0.50$, it is clear that, although uncorrelated in the long run, in the short term there is a decaying antipersistant oscillation,
the origin of which can be clearly tracked to the cyclical behavior of the system caused by a large fraction of contrarians constantly seeking
to occupy the instantaneous global minority state.

\begin{figure}[h!]
\vspace{0.3cm}
    \centering
    \includegraphics[width=0.38\textwidth]{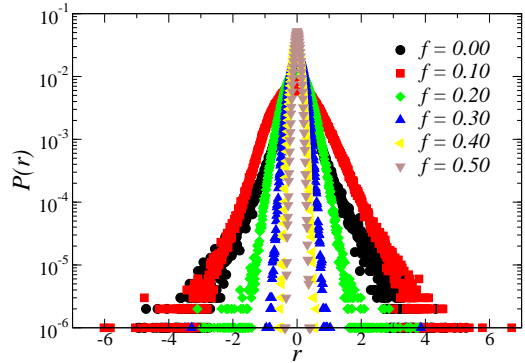} 
    \caption{Distributions of logarithmic returns for a noise parameter $q=0.16$ and various concentrations of contrarians:
    $f=0.00,0.10,0.20,0.30,0.40,0.50$.
    }
   \label{fig:hisret}
\end{figure}

We now proceed to study the distribution of returns of the time series generated by the 3 state model.
As noted before, Fig. \ref{fig:sample} clearly illustrates the decrease of the magnitude of the fluctuations of the 
returns as the number of contrarians is increased. This effect is captured by the tails of the return
distributions, as shown in Fig. \ref{fig:hisret}. 
It is immediately apparent that an increase in the fraction of contrarians produces a reduction in the heaviness of the tails,
greater fluctuations in the returns becoming less likely.
In effect, for high values of $f$, the distribution loses its fat tails and becomes Gaussian, in accordance with the results
of the overly simplistic and widespread model of financial markets given by \textit{Geometric Brownian Motion}, used typically in
Black-Scholes calculations of derivative pricing \cite{Voit}.\\ \\

\begin{figure}[h]
\vspace{0.3cm}
    \centering
    \includegraphics[width=0.4\textwidth]{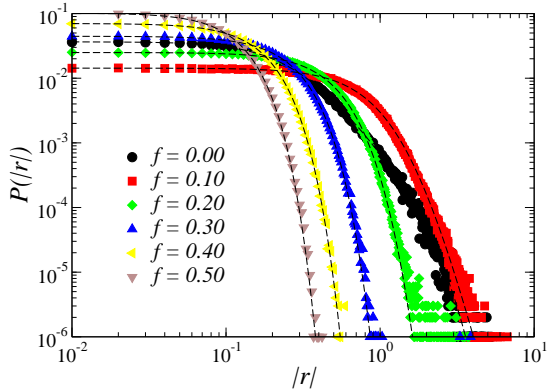} 
    \caption{Distributions of volatility for a noise parameter $q=0.16$ and various concentrations of contrarians: 
    $f=0.00,0.10,0.20,0.30,0.40,0.50$. Fits of the curves to symmetric coupled exponential distributions defined in Eq. (\ref{eq:hisvol}) are presented
    in dashed lines.
    }
   \label{fig:hisvol}
\end{figure}

In order to quantify the transition of the return distributions from a heavy-tailed leptokurtic regime into a Gaussian mesokurtic
regime and, possibly, into a compact-support platykurtic regime, 
we consider distributions from the symmetric \textit{coupled exponential family} \cite{Nelson2017, Nelson2019}. 
Such family of distributions is defined by:
\begin{equation}
    \label{eq:hisvol}
    P_{\sigma,\kappa,\alpha}(r) \equiv \left[ Z(\sigma,\kappa,\alpha)\left(1+\kappa \Bigl|\frac{r}{\sigma} \Bigr|^\alpha \right)^{\frac{1+\kappa}{\alpha \kappa}}_{+} \right]^{-1}
\end{equation}
where $\sigma$, $\kappa$ and $\alpha$ are the parameters of the function and $(a)_+\equiv \max (0,a)$.
We shall refer to the shape parameter $\kappa$ as the \textit{nonlinear statistical coupling} 
and to $\sigma$ as the \textit{scale paramater}
in an interpretation
of nonextensive statistical mechanics \cite{Tsallis2009} which has been used as a model of financial markets \cite{Tsallis2003, Biondo2015} and other
complex systems.

This family of distributions is characterized by the fact that it is capable of capturing the aforementioned transition in a very
natural way. Indeed, when $\kappa>0$, the function exhibits a heavy-tail decay. 
When $\kappa=0$, the function is a generalized Gaussian distribution. When $-1<\kappa<0$, the function is a distribution with compact support.
Moreover, if $\alpha=2$, the function is a coupled Gaussian distribution: with $\kappa=0$, the Gaussian distribution; and with $\kappa>0$, the Student's t distribution with the degree of freedom being the reciprocal of 
$\kappa$, $\nu=1/\kappa$. Therefore, the nonlinear statistical coupling is capable of providing a numerical measure of the transition from a leptokurtic to a platykurtic state.

We fix $\alpha=2$ and use the coupled Gaussian distribution to fit the histograms of the absolute values of the logarithmic returns, which measure 
the volatility of the time series, and focus on the behavior of the nonlinear coupling parameter as a function of the number of contrarians.
With this prescription, the heavy tails of the coupled Gaussian distribution behave as $P_{\sigma,\kappa,2}(r) \sim |r|^{-(1+\kappa)/\kappa} \sim |r|^{-(1+\nu)}$.
Fig. \ref{fig:hisvol} shows the distribution of the absolute values of the logarithmic returns for different concentrations of contrarians
and its fit to the function defined by Eq. (\ref{eq:hisvol}).\\

\begin{figure}[h]
\vspace{0.3cm}
    \centering
    \includegraphics[width=0.4\textwidth]{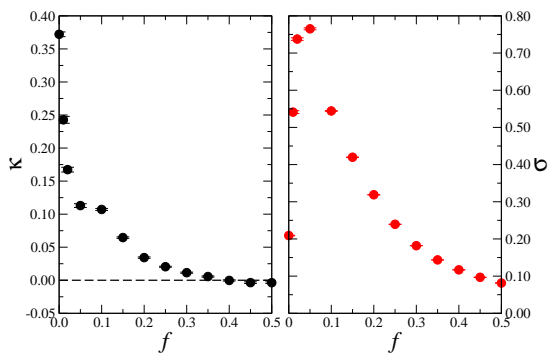} 
    \caption{Coupling and scale parameters of the return distributions for a noise parameter $q=0.16$ and various concentrations of contrarians: 
    $f=0.00, 0.02, 0.05, 0.10, 0.15, 0.20, 0.25, 0.30, 0.35, 0.40, 0.45, 0.50$.
    A dashed horizontal line marks the value of the nonlinear coupling of a Gaussian distribution $\kappa=0$.
    }
   \label{fig:sk}
\end{figure}

In Fig. \ref{fig:sk}, we present the values of the nonlinear coupling parameter $\kappa$ and the scale $\sigma$ of these fits.
Notice that the nonlinear coupling parameter $\kappa$ of the coupled Gaussian distribution monotonically decreases with $f$, 
providing numerical evidence for the progressive loss of the heavy tails as the distribution approaches a Gaussian for high 
values of $f$. In fact, for values of $f\geq 0.4$, $\kappa\approx0$, indicating that the distribution is approximately Gaussian. Further, $\kappa$, although approximately zero,
attains negative values for a higher fraction of contrarians, 
suggesting that the dynamics may indeed produce distributions of compact support in the antipersistent oscillatory regime
present with very high fractions of contrarians.

Therefore, the nonlinear statistical coupling is capable of capturing the complexity of the behavior of the system, numerically measuring
the change from a heavy tailed regime to a non-heavy tailed regime.

The scale parameter $\sigma$, which is a generalization of the standard deviation within the context of the coupled exponential family, decreases for values of $f>0.05$. This is consistent with the increase in the fraction of contrarian traders reducing the variation in the log returns.  
So the contrarian traders decrease variation by reducing both the scale and the shape or coupling of the log-return distribution.
However, it is clear that, for $f<0.05$, the broadness of the distribution has a remarkably
different behavior. This can be qualitatively appreciated in Figs. \ref{fig:hisret} and \ref{fig:hisvol},
where a visual examination of the $f=0.00$ curve relative to the rest indeed suggests a heavier tail exponent but
a smaller scale, which is approximately the knee of the log-log plot of the distributions. This suggests that near $f=0.05$, where $\sigma$ peaks, there is a change 
in the behavior of the scale of the distribution, even though the statistical coupling is strictly monotonically decreasing
with $f$.\\

\begin{figure}[h]
\vspace{0.3cm}
    \centering
    \includegraphics[width=0.38\textwidth]{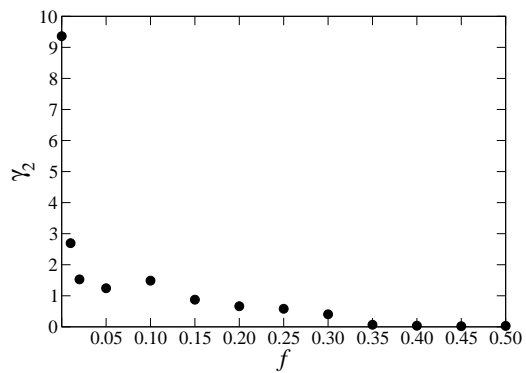} 
    \caption{Excess kurtosis of the return distributions for a noise parameter $q=0.16$ and various concentrations of contrarians. 
    A dashed horizontal line marks the value of the kurtosis of a Gaussian distribution $\gamma_2=0$.
    }
   \label{fig:kurt}
\end{figure}

Further confirmation of the transition from a leptokurtic to a mesokurtic regime is provided by the behavior of the excess kurtosis, 
$\gamma_2\equiv \langle r^4\rangle /\langle r^2\rangle^2 - 3$, of the
return distribution itself, as shown in Fig. \ref{fig:kurt}, as a measure of the heaviness of its tail.
Recall that the excess kurtosis of a Gaussian distribution is $\gamma_2=0$.
It is clear from the figure that an increasing concentration of contrarians tends to remove the heavy tails of the distribution, driving
the transition into a Gaussian regime for high values of $f$.

\begin{figure}[h]
\vspace{0.3cm}
    \centering
    \includegraphics[width=0.4\textwidth]{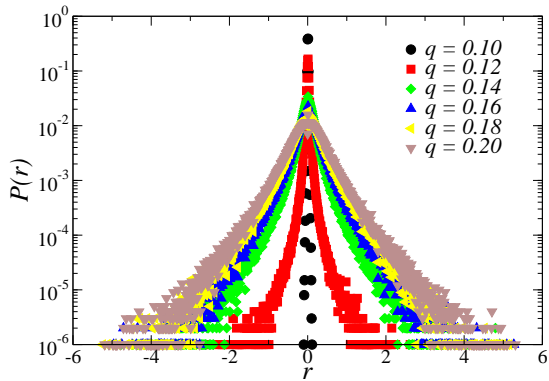} 
    \caption{Distributions of logarithmic returns in the absence of contrarians $f=0.00$ for $q=0.10,0.12,0.14,0.16,0.18,0.20$.
    }
   \label{fig:hisretf000}
\end{figure}

Fig. \ref{fig:hisretf000} shows the distribution of logarithmic returns for different values of $q$ in the absence of contrarians.
An interesting feature of the MVM3 dynamics is that, even in the absence of contrarians ($f=0.00$), the relative sizes in the fluctuations of the 
order parameter as representations of the returns of an asset can be controlled by the noise parameter $q$ alone.
As is to be expected, in an ordered regime with $q<<q_c$, the fluctuations in the spontaneous magnetization of the system are very small,
therefore leading to a state of very small returns and volatility. 
Proximity to the critical point, where fluctuations in the magnetization diverge in the thermodynamic limit, suggests a transition from 
a low variance regime to a more turbulent state with large fluctuations, as can be seen in the figure.

\begin{figure*}[ht]
\vspace{0.3cm}
    \centering
    \subfloat[Subfigure 1 list of figures text][]
        {
        \includegraphics[width=0.4\textwidth]{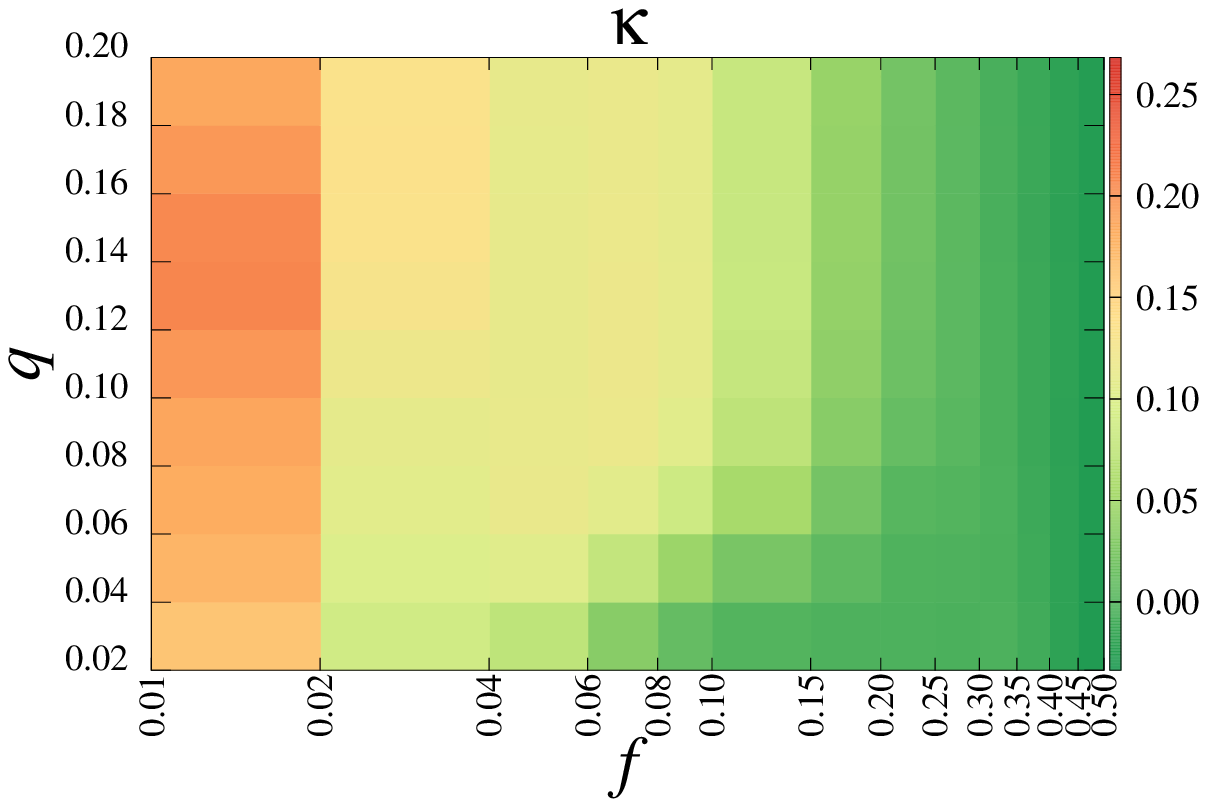}
        \label{fig:heatk}
        }
        \qquad
    \subfloat[Subfigure 2 list of figures text][]
        {
        \includegraphics[width=0.4\textwidth]{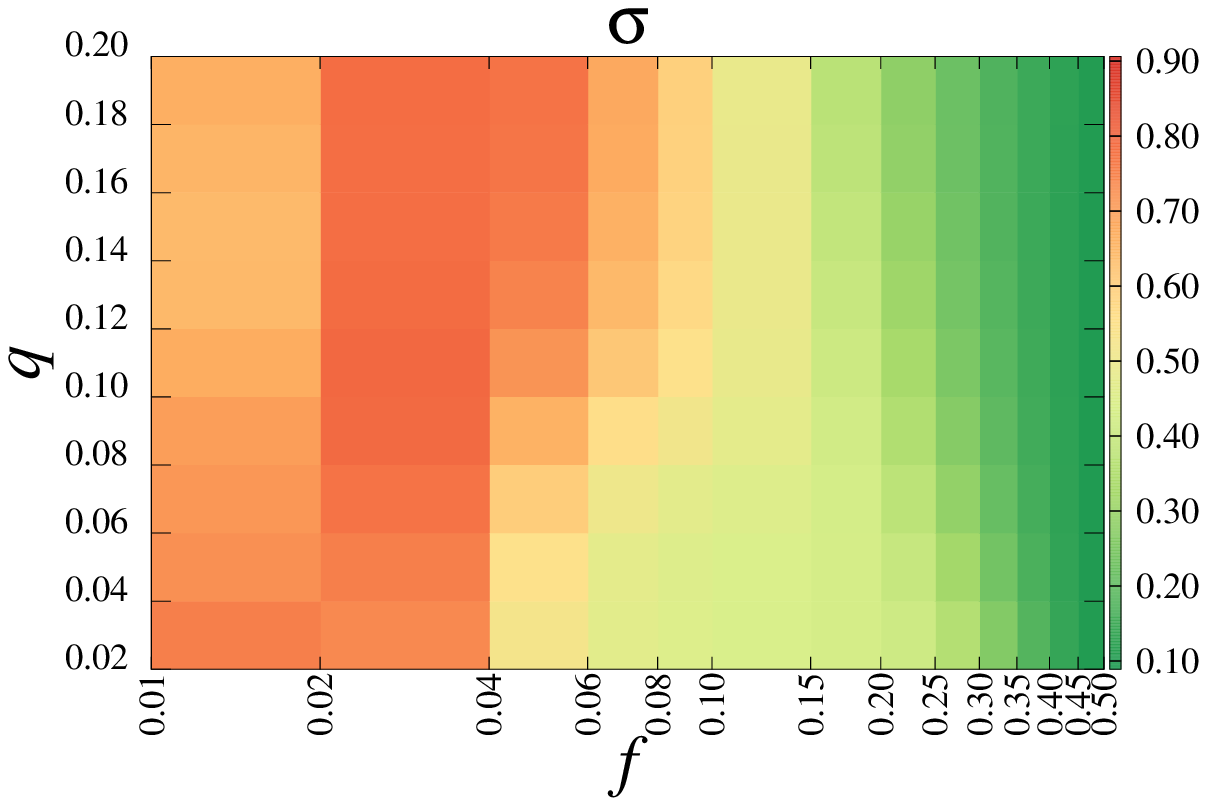}
        \label{fig:heats}
        }
    \caption{Heat map of (a) the nonlinear statistical coupling $\kappa$ and (b) the scale parameter $\sigma$ as a function of the noise parameter $q$ and the percentage of contrarians $f$.}
    \label{fig:heatmaps}
\end{figure*}

A more comprehensive picture of the behavior of the model emerges when looking at Fig. \ref{fig:heatmaps},
in which we present the heat maps of the fitting parameters
$\kappa$ and $\sigma$ for different pairs of values $(q,f)$ of the model's parameters.
A first feature that one can extract from the heat map of the nonlinear statistical coupling is the monotonic
decreasing of $\kappa$, and therefore the transition from a leptokurtic regime to a platikurtic regime, 
as one increases the fraction of contrarians for any fixed value of the noise parameter, $q$. Indeed, $\kappa$ becomes negative for high enough values of $f$,
suggesting that the distributions not only lose their fat tails as they approach a Gaussian regime,
they also eventually transition into a state where they exhibit compact supports ($\kappa<0$),
as discussed previously for Fig. \ref{fig:sk}. It is clear from the heat map that the transition,
for fixed $q$, from heavy tails to a complete loss of tails with increasing fraction of contrarians occurs 
faster along the $f$ axis for smaller values of $q$.
These observations suggest that the nonlinear statistical coupling is capable of capturing the degree of 
complexity of the system, as reflected by behavior of the tails of the distributions.

The heat map of the scale parameter is also consistent with our discussion of Fig. \ref{fig:sk},
showing, for fixed values of $q$, that $\sigma$ increases rapidly for small values of $f$, reaches a maximum,
and consequently decreases for large fractions of contrarians. So even though for small values of $f$ 
the scale parameter (as a measure of the distribution's width) is small, the distributions still exhibit heavy tails,
as evidenced by the nonlinear coupling $\kappa$. 
Notice that the scale parameters seem to consistently peak around the same values of $f$ for fixed values of $q$.
However, the peak of this transition in the scale parameter is wider for small values of $q$
than it is for high values of $q$.

For the smallest value of $f$ recorded ($f=0.01$), one can also appreciate the fact that, for very small 
values of the noise parameter ($q \leq 0.01$), $\sigma$ increases monotonically with $q$.
This is suggestive of the fact that the spread or variance of the distribution actually becomes smaller with increasing
values of $f$ for fixed $q$. This is the effect previously seen in the Fig. \ref{fig:hisretf000}, where the 
spread of the distribution grows noticeably with increasing $q$.
But this behavior of the scale parameter (for fixed $q$) is actually reversed for values of $f\geq0.02$, 
as seen in the heat map, $\sigma$ then increasing monotonically with $q$ for fixed $f$, suggesting a change of regime
in the behavior of the scale parameter around $f\approx 0.02$. 
So, for any fixed macroscopically relevant concentration of contrarians ($f\gtrsim 0.02$), the scale paramater increases with
the social temperature $q$. 

\section{CONCLUSION}
\label{sec:conclusion}

In this work, we propose a three-state, agent-based, microscopic market model with stochastic dynamics.
It features a heterogenous population of traders comprised of two categories: noise traders and contrarians.
Noise traders interact locally with their nearest neighbors, tending to agree with the state of the local majority.
Contrarians are subject to global interactions with the market as a whole and they tend to follow the state of the global minority.
By relating changes in the order parameter of this system to price fluctuations in the market, the simulation of the dynamics
of the model is capable of reproducing the main qualitative and quantitative features of real financial time series, such
distributions of returns with long tails, volatility with long-term memory and volatility clustering.

The logarithmic returns of the simulation fit a coupled Gaussian distribution, which is parameterized by the scale or generalized standard deviation 
and the shape or nonlinear statistical coupling.  
For macroscopically relevant fractions of contrarians ($f>0.02$), an increase in the contrarians decreases both the scale and the shape of the distribution. 
In this same region, increasing the probability of dissent $q$ also increases the scale and shape of the returns. 
For a very small fraction of contrarians ($f<0.02$), the behavior of the scale is more complex, with a maximum being reached and then decreasing 
as the percentage of contrarians decreases.

The model's simplicity is capable of shedding light into the potential mechanisms at play behind the social psychology of decision making
and opinion formation in a financial market.

\center{\textbf{Acknowledgements}}

The authors acknowledge financial support from Brazilian and Chinese institutions and funding agents UPE (PFA2018,
PIAEXT2018), FACEPE (APQ-0565-1.05/14, APQ-­0707­-1.05/14), CAPES, CNPq, National Natural Science Foundation of China (61603011),
the International Postdoctoral Exchange Fellowship Program (20170016) and Beijing Social Science Foundation of China
(16JDGLC005). The Boston University Center for Polymer Studies is supported by NSF Grants PHY-1505000, CMMI-1125290,
and CHE-1213217, by DTRA Grant HDTRA1-14-1-0017, and by DOE Contract DE-AC07-05Id14517.

\bibliography{MVMPAPER.bib}
\newpage 
\end{document}